\documentclass{ws-mpla}
\usepackage[super]{cite}
\usepackage{graphicx}
\usepackage[a4paper,left=30mm,right=30mm,top=43mm,bottom=25mm]{geometry}
\usepackage{relsize}

\usepackage{color}
\definecolor{coolblack}{rgb}{0.0, 0.18, 0.39}

\usepackage[hidelinks,bookmarks=true,
colorlinks = true,
linkcolor = red,
urlcolor  = coolblack,
citecolor = blue,
anchorcolor = blue]{hyperref}

\usepackage[numbered]{bookmark}

\begin{document}

\relscale{1.31} 

\markboth{{\relscale{1.3}Ahmad Mohamadnejad}}
{{\relscale{1.3}Cosmological evolution of the proton-to-electron mass ratio ...}}

\catchline{}{}{}{}{}

\title{{\relscale{1.3}Cosmological evolution of the proton-to-electron mass ratio in an extended Brans-Dicke theory}}

\author{\footnotesize {\relscale{1.3}Ahmad Mohamadnejad}}

\address{{\relscale{1.3}Young Researchers and Elite Club, Islamshahr Branch, Islamic Azad University, Islamshahr 3314767653, Iran\\
a.mohamadnejad@ut.ac.ir}}
\date{\today}

\maketitle

\pub{}{\today}

\begin{abstract}
{\relscale{1.25}We study variation of the proton-to-electron mass ratio $ \mu=m_{p}/m_{e} $ by incorporating standard model (SM) of particle physics into an extended
Brans-Dicke theory. We show that the evolution of the Higgs vacuum expectation value (VEV), with expansion of the Universe, leads to the variation of the proton-to-electron mass ratio. This is because the electron mass is proportional to the Higgs VEV, while the proton mass is mainly dependent on the quantum chromodynamics (QCD) energy scale, i.e., $ \Lambda_{QCD} $. Therefore, using the experimental and cosmological constraints on the variation of the $ \mu $ we can constrain the variation of the Higgs VEV. This study is important in understanding the recent claims of the detection of a variation of the proton-to-electron mass ratio in quasar absorption spectra.}

\end{abstract}


\pdfbookmark[section]{Introduction}{sec1}
\section{Introduction}
The assumption of the constancy of the constants plays an important role in astronomy and cosmology, especially with respect to the look-back time measured by the redshift.
Refusing the possibility of varying constants could lead to a distorted view of the Universe and, if such a variation were founded, corrections would have to be applied.
Therefore, it is important to investigate this possibility, in particular as the measurements become more and more strict.
The history of these investigations traces back to early ideas in the 1930s on the possibility of a time variation of the gravitational constant $ G $ by Milne \cite{Milne} and the suggestion by Dirac of the large number hypothesis \cite{Dirac:1937ti,Dirac:1938mt} which led him also to propose in 1937 the time evolution of $ G $. 
Clearly, the constants have not undergone enormous variations on solar system scales and geological time scales, and one is looking for tiny effects. There are some reviews on the time variation of the constants of nature, for example see \cite{Uzan:2002vq,Uzan:2010pm,Chiba:2011bz,Calmet:2014qxa}.

The proton-to-electron mass ratio $ \mu \equiv \frac{m_{p}}{m_{e}} $ is another important dimensionless constant that may evolve with cosmic time. This ratio is effectively the ratio of the strong and electroweak scales.
The molecular absorption lines spectra, as first pointed out by Thompson \cite{Thompson}, can provide a test of the variation of $ \mu $.
The variation of this ratio has been measured using quasar absorption lines \cite{Reinhold:2006zn,Flambaum:2007fa,King:2011km,Bagdonaite:2011ab,Rahmani:2013dia,Dapra:2017vlu}.
Motivated by these observations of quasar absorption systems, some theories have been proposed, for example see \cite{Barrow:2005qf,Calmet:2006sc,Dent:2006fn,Chiba:2006xx,Lee:2007du,Avelino:2008dc,Calmet:2017czo,Fritzsch:2016ewd}.
In this paper, considering the standard model of particle physics and a generalized Brans-Dicke (BD) theory containing two interacting scalar fields, we propose another scenario in which the proton-to-electron mass ratio varies with cosmic time.

According to the standard model of particle physics, the masses of elementary particles, including the electron mass, is proportional to the vacuum expectation value (VEV) of the Higgs boson while the proton mass is mainly determined by the scale of quantum chromodynamics (QCD), i.e., $ \Lambda_{QCD} $ \cite{Gasser:1982ap}. Therefore, the variation of the Higgs VEV, $ \nu $, leads to a variation of the electron mass $ m_{e} $ while the proton mass $ m_{p} $ almost remains constant (We have also supposed that Yukawa coupling constant between Higgs boson and electron is time-independent). On the other hand, there must be a time, for example the time of electroweak phase transition, when Higgs VEV were different from what is today. Therefore, it is natural to expect that Higgs VEV has been varied during cosmological epochs. Indeed, in the framework of generalized Brans-Dicke (BD) theories, in which BD-field $ \psi $ interacts with Higgs field $ \varphi $, one can show the effective potential associated to $ \varphi $ may adopt
the Higgs potential form with a slowly evolving vacuum expectation value.

In the next section we discuss the variation of the electron mass and proton mass with the evolution of the Higss VEV. The study of the evolution of the Higgs VEV in cosmological context comes next. Finally, in the conclusion section, using experimental and cosmological observations, we put a constraint on the variation of the Higss VEV.

\pdfbookmark[section]{Standard Model and electron/proton mass}{sec2}
\section{Standard Model and electron/proton mass}
The Higgs boson is presumed by the electroweak theory initially  to explain the origin of particle masses. In the Standard Model of particle physics, the mass of all elementary particles is proportional to the Higgs VEV. Therefore, for the electron mass as well as the quarks masses, we have 
\begin{equation}
m_{e,q} = \lambda_{e,q} \nu , \label{2-1}
\end{equation}
where $ \lambda_{e,q} $ is the Yukawa coupling and $ \nu $ is the Higgs VEV. On the other hand, how the proton mass is related to the Higgs VEV is another question. The total proton
mass is 938 MeV, while the masses of the valence quarks in the proton are just $ \sim $3 MeV per quark which is
directly related to the Higgs VEV. The quark and gluon contributions to the proton mass can be provided
by solving QCD non-perturbatively. The proton mass $ m_{p} $ is a function of $ \Lambda_{QCD} $ and quark masses.
In the chiral limit the theory contains massless physical particles. 
In this limit $ m_{p} $ is proportional to the QCD
energy scale $ \Lambda_{QCD} $ and it does not change with variation of the Higgs VEV. Considering quark masses, the quark mass expansion for the proton mass has the structure \cite{Gasser:1982ap}
\begin{equation}
 m_{p} = a(\Lambda_{QCD}) + \sum\limits_{q} b_{q} m_{q} . \label{2-2}
\end{equation}
One can derive a separation of
the nucleon mass into the contributions from the quark,
antiquark, gluon kinetic and potential energies, quark
masses, and the trace anomaly \cite{Ji:1994av}. The QCD Hamiltonian can be separated into four gauge-invariant parts
\begin{equation}
 H_{QCD} = H_{q} + H_{m} + H_{g} + H_{a} , \label{2-3}
\end{equation}
with
\begin{align}
H_{q} &= \int d^{3} x \, \psi^{\dagger} (D . \gamma) \psi \label{2-4} , \\
H_{m} &= \int d^{3} x \, \psi^{\dagger} m \psi , \label{2-5} \\
H_{g} &= \int d^{3} x \, \frac{1}{2}(\textbf{E}^{2} + \textbf{B}^{2}) , \label{2-6} \\
H_{a} &= \int d^{3} x \, \frac{9 \alpha_{s}}{16 \pi}(\textbf{E}^{2} + \textbf{B}^{2}) , \label{2-7}
\end{align}
where $ \psi $ is the quark field, and $ \textbf{E} $ and $ \textbf{B} $ are electric and magnetic gluon field strengths, respectively. In Eq. (\ref{2-3}), $ H_{q} $ represents the quark and antiquark kinetic and potential energies, $ H_{m} $ is the quark mass term, $ H_{g} $ represents the gluon energy, and finally, $ H_{a} $ is the trace anomaly term. Recent lattice simulations shows that the joint u/d/s quark mass term, $ H_{m} $, only contribute 9 percent of the proton mass \cite{Yang:2018nqn}. Note that this term is the one where the Higgs boson contributes and it means the second term in Eq. (\ref{2-2}) contribute 9 percent of the proton mass.
Following \cite{Calmet:2017czo}, we assume that $ \Lambda_{QCD} $ as well as Yukawa couplings are constants, while the Higgs VEV $ \nu $ can evolve.
Note that it is not the only option. For example in \cite{Fritzsch:2012qc} such assumption is not made and nevertheless an alternative scenario changing the proton-to-electron mass ratio is obtained.

According to Eqs. (\ref{2-1}) and (\ref{2-2}) we have
\begin{align}
\frac{\Delta m_{e}}{m_{e}} &= \frac{\Delta \nu}{\nu} , \label{2-8} \\
\frac{\Delta m_{p}}{m_{p}} &= \frac{\sum\limits_{q} b_{q} m_{q}}{a(\Lambda_{QCD}) + \sum\limits_{q} b_{q} m_{q}} \frac{\Delta \nu}{\nu} = \frac{9}{100} \frac{\Delta \nu}{\nu} . \label{2-9}
\end{align}
In Eq. (\ref{2-9}), as we discussed before, we have assumed the  quark mass term, $ \sum\limits_{q} b_{q} m_{q} $, contribute 9 percent of the proton mass. Thus $ \frac{\Delta m_{p}}{m_{p}} $ is negligible in comparison with $ \frac{\Delta m_{e}}{m_{e}} $ and it means, considering time-dependent Higgs VEV, the electron mass should change with the cosmic time while the proton mass remains (almost) constant.

According to Eqs. (\ref{2-8}) and (\ref{2-9}), the cosmological evolution of the Higgs VEV would lead to a cosmological variation of the proton-to-electron mass ratio $ \mu $:
\begin{equation}
\frac{\Delta \mu}{\mu} = \frac{\Delta m_{p}}{m_{p}} - \frac{\Delta m_{e}}{m_{e}} =  - \frac{91}{100} \frac{\Delta \nu}{\nu}, \label{2-10}
\end{equation}
where $ \Delta \nu / \nu \equiv (\nu_{z}-\nu_{0}) / \nu_{0} $ for $ \nu_{0} $ and $ \nu_{z} $ the present value of the Higgs VEV and its value in the absorption cloud at redshift $ z $, respectively.

In order to provide another estimation for $ \frac{\Delta m_{p}}{m_{p}} $ as function of $ \frac{\Delta \nu}{\nu} $, we consider the relation between top quark mass and proton mass discussed in \cite{Quigg:1996ew}. In \cite{Quigg:1996ew}, using the
renormalization group equation, it is argued that in a simple unified theory the proton mass depends on the top quark mass, $ m_{t} $, as the following
\begin{equation}
\frac{m_{p}}{1 \, GeV} \propto \left( \frac{ m_{t}}{1 \, GeV} \right) ^{\frac{2}{27}}. \label{2-11}
\end{equation}
According to this relation, we get
\begin{equation}
\frac{\Delta m_{p}}{m_{p}} = \frac{2}{27} \frac{\Delta \nu}{\nu} = \frac{7.4}{100} \frac{\Delta \nu}{\nu} , \label{2-12}
\end{equation}
which is in good agreement with Eq. (\ref{2-9}).

In the following, we will show how the Higgs VEV, and consequently the proton-to-electron mass ratio, varies with cosmic time.

\pdfbookmark[section]{Expansion of the Universe and evolution of the Higgs VEV}{sec3}
\section{Expansion of the Universe and evolution of the Higgs VEV}
Consider the Higgs sector of the standard model
(in unitary gauge) with the following Lagrangaian
\begin{equation}
L =  - \frac{1}{2} \partial_{\mu} \varphi  \partial^{\mu} \varphi - \frac{\lambda}{4} (\varphi^{2} -\nu^{2})^{2},
 \label{005}
\end{equation}
where $ \nu_{0} = 246 \, \, GeV $ is the present Higgs VEV.

One can decompose $ \varphi $ to the Higgs particle $ h $ and a classical background field $ \phi $ which plays the role of a Higgs cosmological value depending on the cosmic time: $ \varphi = \phi(t) + h $. In Lagrangian (\ref{005}), according to spontaneous symmetry breaking, we should replace $ \varphi $ with $ \nu_{0} + \phi(t) + h = \nu(t) + h  $ where the time-dependent Higgs vacuum expectation value defined as $ \nu(t) \equiv \nu_{0} + \phi(t) $. Then by minimally coupling $ \phi(t) $ to gravity, one can show that the expansion of the Universe leads to a cosmological time evolution of the vacuum expectation of the Higgs boson \cite{Calmet:2017czo}.
Time-dependent Higgs VEV is also studied in \cite{Casadio:2007ip} considering dynamics of the standard model of particle physics alone. In this work, it has been shown that variation of the Higgs VEV leads to the non-adiabatic production of both bosons and fermions. There are other motivations for time-dependent Higgs VEV as well. For example in \cite{Sola:2016our}, the structure of the Higgs potential has been derived from a generalized Brans-Dicke theory \cite{Brans:1961sx} containing two interacting scalar fields.
By requiring that the cosmological solutions of the
model are consistent with observations, it has been shown that the effective scalar field potential adopts the Higgs potential form with a mildly time-dependent Higgs VEV. We conclude that time-dependent Higgs VEV is a natural assumption specially in the context of gravity and cosmology.

Motivated by these arguments, here we assume the Higgs VEV could be time-dependent. However, this does not mean that it certainly varies with time. We just consider it as a well-motivated possibility. Therefore, keeping an open mind on this possibility, we let equations determine the dynamics of its variation.
The most stringent bound on $ \frac{\dot{\nu}_{0}}{\nu_{0}} \simeq - \frac{\dot{\mu}_{0}}{\mu_{0}} $ comes from the comparison of the transitions in Yb$^{+} $ with the cesium atomic clock \cite{Huntemann:2014dya}:
\begin{equation}
\frac{\dot{\mu}_{0}}{\mu_{0}} = (-0.5 \pm 1.6)\times 10^{-16} \, \, year^{-1} .
 \label{017}
\end{equation}
Comparing this value with Hubble constant $ H_{0} \simeq 7 \times 10^{-11}  \, \, year^{-1} $, we obtain an experimental limit $ \frac{\dot{\nu}_{0}}{\nu_{0}} \simeq 10^{-6} H_{0} $.
In the end, one can constrain higgs VEV variation using this experimental data and cosmological observations coming from quasar absorption
spectra.

In this section, we consider a generalized Brans-Dicke theory containing two interacting scalar fields used in \cite{Sola:2016our}. In this approach, a mildly time-dependent Higgs VEV is derived from evolution of another scalar field.
Note that Brans-Dicke theories are connected to the running vacuum models \cite{Peracaula:2018dkg,Perez:2018qgw}. These models can also produce a change of the vacuum with the expansion of the universe and they are highly competitive in the fitting of the cosmological data, see e.g. \cite{Gomez-Valent:2018nib,Sola:2017jbl,Sola:2016ecz,Sola:2016jky}

The Universe is isotropic and homogeneous, at least as a first approximation. This assumption will lead us to the Robertson-Walker metric.
Almost all of modern cosmology is based on this metric:
\begin{equation}
ds^{2} = - d t^{2} + a^{2}(t) \left[ \frac{d r^{2}}{1 - K r^{2}} + r^{2} d \Omega \right] , \label{002}
\end{equation}
where $ K $ is the curvature signature and $ a(t) $ is the scale factor.
Now consider an extend Brans-Dicke model with the BD-field, $ \psi $, and another scalar field $ \varphi $ which eventually plays the role of Higgs field. There is a coupling term between these scalar fields, as well as non-minimal interactions with gravity. The new Brans-Dicke model includes all of the terms of the SM, however, we only consider the part of this action containing the dynamics and interaction between the scalar fields, as well as their couplings to gravity. This action is used in \cite{Sola:2016our} in which the authors proposed a physical motivation for the structure of the Higgs potential.
The scalar part of the action is:
\begin{align}
I \supset&  \int d^{4} x \sqrt{- g} ( \frac{1}{2} R \psi - \frac{\omega}{2 \psi} g^{\mu \nu} \partial_{\mu} \psi  \partial_{\nu} \psi -\frac{1}{2} g^{\mu \nu} \partial_{\mu} \varphi  \partial_{\nu} \varphi - V(\varphi) \nonumber \\ 
&+ \eta \varphi^{2} \psi + \xi R \varphi^{2} + \frac{1}{\varphi^{2}} S_{\mu \nu} \partial^{\mu} \varphi  \partial^{\nu} \varphi ) ,
 \label{321}
\end{align}
where $ R $ is Ricci scalar of curvature, and $ \omega $, $ \eta $, and $ \xi $ are dimensionless constants. In action (\ref{321}), $ S_{\mu \nu} $ is
\begin{equation}
S_{\mu \nu} =  \varsigma R_{\mu \nu}- \frac{\theta}{2} g_{\mu \nu} R,
 \label{322}
\end{equation}
which allows generalized derivative couplings of $ \varphi $ with gravity. The couplings $ \varsigma $ and $ \theta $ in (\ref{322}) were first studied in \cite{Amendola:1993uh,Capozziello:1999uwa}.
So far, the Higgs potential $ V(\varphi) $ in (\ref{321}) is unidentified. However, it takes the Higgs potential form after
imposing the appropriate conditions.

The Euler-Lagrange equations corresponding to action (\ref{321}) in the cosmological context with Robertson-Walker metric, Eq. (\ref{002}), are given by \cite{Sola:2016our}
\begin{align}
&V(\varphi) ={{3}H^{2}\psi}+{{3}H{\dot{\psi}}}-\frac{\omega}{2}\frac{\dot{\psi}^{2}}{\psi}+\eta \varphi^{2}\psi-\frac{1}{2}\dot{\varphi}^{2} +6\xi H^{2}\varphi^{2} \nonumber \\
& +12\xi H \dot{\varphi}\varphi  -{9}\theta H^{2}\frac{\dot{\varphi}^{2}}{\varphi^{2}}- 6\left(\theta-\varsigma\right)\dot{H}\frac{\dot{\varphi}^{2}}{\varphi^{2}}+6\left(\theta-\varsigma\right)H\frac{\dot{\varphi}\ddot{\varphi}}{\varphi^{2}} \nonumber \\
&-6\left(\theta-\varsigma\right){H}\frac{\dot{\varphi}^{3}}{\varphi^{3}}\,, \ \ \ \ \label{323}
\end{align}
\begin{align}
&\ddot{\varphi}+3 H \dot{\varphi}-12\xi\dot{H}\varphi - 24\xi H^{2}\varphi+\frac{d V}{d\varphi}+6 \left(2\theta -\varsigma\right) H^{2}\left(\frac{\ddot{\varphi}}{\varphi^{2}}-\frac{\dot{\varphi}^{2}}{\varphi^{3}}\right) \nonumber \\
&+18 \left( 2 \theta -\varsigma \right) {H}^{3}\frac{\dot{\varphi}}{\varphi^{2}}+
6 \left(7 \theta -5 \varsigma \right) H \dot{H}\frac{\dot{\varphi}}{\varphi^{2}}+6 \left(\theta - \varsigma \right)\ddot{H}\frac{\dot{\varphi}}{\varphi^{2}} \nonumber\\
&+6 \left(\theta -\varsigma\right)\dot{H}\left(\frac{\ddot{\varphi}}{\varphi^{2}}-\frac{\dot{\varphi}^{2}}{\varphi^{3}}\right)-{2}\eta \psi \varphi= 0\,, \label{324}
\end{align}
\begin{equation}
3\dot{H}+6{H}^{2} - \omega \frac{\ddot{\psi}}{\psi}+\frac{ \omega}{2} \frac{\dot{\psi}^{2}}{{\psi}^{2}}-3 H \omega\frac{\dot{\psi}}{\psi}+ \eta {\varphi} ^{2}=0\,,  \ \ \ \ \ \ \ \ \ \ \label{325}
\end{equation}
where $ H=\frac{\dot{a}}{a} $ is the Hubble parameter.

The above equations can take the power-law solutions:
\begin{equation}
\psi (t) \propto t^{\alpha}, \quad  \varphi (t) \propto t^{\beta}, \quad H \propto t^{-1} ,
 \label{326}
\end{equation}
where $ \alpha $ and $ \beta $ are dimensionless parameters. Since $ \alpha $ controls the evolution of Newton's gravitational coupling in the context
of BD-model, we expect $ \alpha $ to be small. Indeed, from observation we know that $ \vert \alpha \vert \leq 10^{-3} $ \cite{Uzan:2010pm}, but its sign is undetermined, see e.g. \cite{Li:2013nwa}. Furthermore, consistency of Euler-Lagrange equation (\ref{325}) with power-law solutions (\ref{326}) implies $ \beta = -1 $. 

According to the solution (\ref{326}), the first three terms of $ V(\varphi) $ in (\ref{323}) are proportional to $ t^{\alpha} \varphi^{2} $ while the other terms are proportional to $ \varphi^{4} $. Therefore, equation (\ref{323}) leads to the Higgs-like potential for field $ \varphi $
\begin{equation}
V (\varphi) = -  \frac{1}{4} M_{\varphi}^{2} (t) \varphi^{2} + \frac{\lambda}{4} \varphi^{4}, 
 \label{327}
\end{equation}
where $ M_{\varphi}^{2} (t) \propto t^{\alpha} $. 
Eqs. (\ref{324}) and (\ref{325}) lead to two constraints for the parameters of extended BD-action and constants in solution (\ref{326}), so they are irrelevant in our study.
In this approach parameters of Higgs potential (\ref{327}) are related to the parameters of extended BD-action (\ref{321}) which can be fixed so that $ M_{\varphi}^{2} (t) $ and $ \lambda $ be positive and take SM values. Actually it is easy to see why $ M_{\varphi}^{2} (t) $ should be positive. The sign of $ M_{\varphi}^{2} (t) $ would be determined by the BD parameter $\omega  $ in the third term of (\ref{323}) and the cosmological constraint $\omega > 890 $ \cite{Avilez:2013dxa}, leads to the negative sign in (\ref{327})).

Regarding Eq. (\ref{327}), Higgs VEV is given by
\begin{equation}
\frac{\partial V}{\partial \varphi} \bigg\rvert_{\nu}=0 \, \Rightarrow \, \nu (t) = \sqrt{\frac{M_{\varphi}^{2} (t)}{2 \lambda}} .  \label{328}
\end{equation}
Therefore, the Higgs VEV is depending on the cosmic time.
\begin{equation}
\nu (t) = \nu_{0} (\frac{t}{t_{0}})^{\alpha / 2} . 
 \label{329}
\end{equation}
As we mentioned before, we have an experimental limit $ \frac{\dot{\nu}_{0}}{\nu_{0}} \simeq 10^{-6} H_{0} $. This limit  leads to $ \alpha \simeq 10^{-5} $. This constraint is much stronger than what we mentioned before, i.e., $ \vert \alpha \vert \leq 10^{-3} $.
However, this bound strongly depends on our assumption, i.e., constant $\Lambda_{QCD}$.

Now, according to Eqs. (\ref{2-10}) and (\ref{329})
\begin{equation}
\frac{\Delta \mu}{\mu} = \frac{91}{100} \left( 1 - (\frac{t}{t_{0}})^{\frac{\alpha}{2}} \right) \simeq  \frac{91}{200} \alpha \ln{\frac{t_{0}}{t}} ,
 \label{330}
\end{equation}
where we have used $ x=1+ \ln{x} $ for $ x \simeq 1 $. Since $ \alpha $ is very small and the formula is logarithmic,  variation of the proton-to-electron mass ratio is very mild.

\pdfbookmark[section]{Results and Conclusion}{sec4}
\section{Results and Conclusion}
\begin{figure}[ht]
\centerline{\includegraphics[scale=0.8]{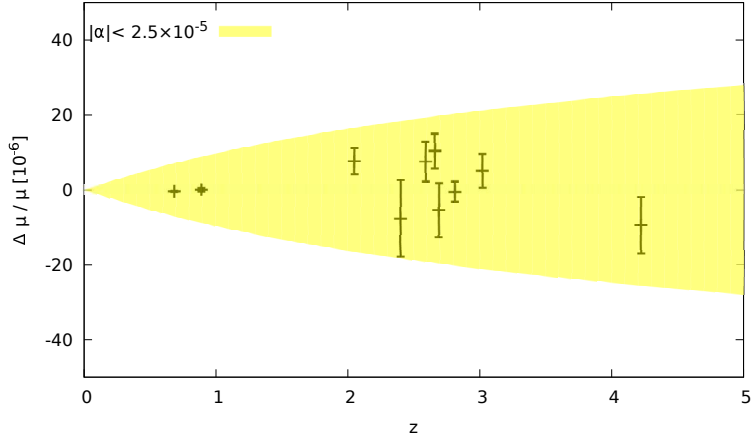}}
\vspace*{8pt}
\caption{Variation of the proton-to-electron mass ratio versus redshif $ z $; there are also some constraints from observations listed in Table \ref{table1}.\protect\label{fig1}}
\end{figure}

In Fig.~\ref{fig1} variation of the proton-to-electron mass ratio $ \Delta \mu / \mu $ versus redshift $ z $ has been depicted for $ |\alpha| < 2.5 \times 10^{-5} $. The colored region is compatible with observational data which can be seen in Fig.~\ref{fig1}. These data are given in Table~\ref{table1} in which the weighted average over observational values for eight quasar $ H_{2} $ absorption spectra \cite{King:2008ud,Malec:2010xv,vanWeerdenburg:2011ru,Wendt:2010qe,Dapra:2016dqh} are taken from \cite{Ubachs:2015fro}).
Apart from the $ H_{2} $ spectra for $ z > 2 $, we also show two observational data of other  molecules spectra for $ z < 1 $ \cite{Murphy:2008yy,Kanekar:2011yb,Henkel:2009cd}.

\begin{table}[ht]
\tbl{Values for $ \Delta \mu / \mu $ obtained for eight $ H_{2} $ quasar absorption spectra at $z > 2$ \cite{Ubachs:2015fro}. For comparison, two results for other molecules spectra at
lower redshift ($z < 1$) are given as well \cite{Murphy:2008yy,Kanekar:2011yb,Henkel:2009cd}.}
{\begin{tabular}{@{}ccc@{}} \toprule
Quasar & Redshift $ z $ & $ \Delta \mu / \mu \, \, [10^{-6}] $ \\ [0.5ex] 
 \hline
B0218+357 & 0.685 & $ -0.35  \pm 0.12 $ \\
PKS1830-211 & 0.89 & $ 0.08  \pm 0.47 $ \\
HE0027-1836 & 2.40 & $ -7.6  \pm 10.2 $ \\
Q0347-383 & 3.02 & $ 5.1  \pm 4.5 $ \\
Q0405-443 & 2.59 & $ 7.5  \pm 5.3 $ \\
Q0528-250 & 2.81 & $ -0.5  \pm 2.7 $ \\
B0642-5038 & 2.66 & $ 10.3  \pm 4.6 $ \\
J1237+064 & 2.69 & $ -5.4  \pm 7.2 $ \\
J1443+2724 & 4.22 & $ -9.5  \pm 7.5 $ \\
J2123-005 & 2.05 & $ 7.6  \pm 3.5 $ \\
 [1ex] \botrule
\end{tabular}\label{table1} }
\end{table}

Note that, the Higgs potential itself has been derived from an extended BD-action with a mildly time-dependent Higgs VEV.
Our result is compatible with BBN constraint on variation of the Higgs VEV, i.e., $ | \Delta \nu / \nu | \lesssim 10^{-2} $ \cite{Yoo:2002vw}. According to Eq. (\ref{330}), for $ t_{0} \sim 13.8 $ Gyr and $ t_{BBN} \sim 3 $ min (with $ | \alpha / 2 | \lesssim 10^{-5} $), we obtain  $ | \Delta \nu / \nu | \lesssim 10^{-4} $ which is compatible with $ | \Delta \nu / \nu | \lesssim 10^{-2} $.

The field $ \psi $ is also related to the effective gravitational "constant" in BD theory, $ G_{eff} \propto \frac{1}{\psi} \propto t^{-\alpha} $, therefore, one can constrain $ \alpha $ using the constraints on the gravitational constant $ G $ \cite{Uzan:2010pm} which, as mentioned before, leads to
$ | \alpha | \lesssim 10^{-3} $.
On the other hand, observations of $ H_{2} $ quasar absorption spectra, including those listed  in Table \ref{table1}, set a constraint on a varying proton-to-electron mass ratio of $ | \frac{\Delta \mu}{\mu} | < 5 \times 10^{-6} \, \, (3-\sigma) $ holding for redshifts in the range $ z = 2.0 - 4.2 $ (for a review see \cite{Ubachs:2015fro}). According to Eq. (\ref{2-10}) this directly constrains the variation of the Higgs VEV or equivalently the parameter $ \alpha $ in (\ref{329}). As it is shown in Fig.~\ref{fig1} we estimate that $ | \alpha | \lesssim 2.5 \times 10^{-5} $.
Note that this bound depends on our particular model in which we have assumed $\Lambda_{QCD}$ is NOT running.

Finally, we should mention that environmental conditions, such as the presence of strong gravitational fields \cite{Bagdonaite:2014mfa}, can also affect the variation of the proton-to-electron mass ratio, and these effects can distort look-back times effects discussed here.

\section*{Acknowledgement}
This work is supported financially by the Young Researchers and Elite Club of Islamshahr Branch of Islamic Azad University.

\end{document}